\documentclass{aa501}
\usepackage{graphicx}

\begin{document}

\title{What can we learn from the surface 
chemical composition of the optical companions of Soft X-ray transients?}

\author{Ene Ergma\inst{1,2} \and Marek J. Sarna\inst{3}}

\offprints{E. Ergma}

\institute{
           Physics Department, Tartu University, {\"U}likooli 18, 50090
           Tartu, Estonia\\
\and
           Astronomical Observatory, Helsinki University, Box 14, 00014 
           Helsinki, Finland\\
           email: ene@physic.ut.ee\\
\and
           N. Copernicus Astronomical Center, Polish Academy of Sciences,
           ul. Bartycka 18, 00--716 Warsaw, Poland.\\ 
           email: sarna@camk.edu.pl\\
}

\date{Received ;accepted}

\abstract{
Several evolutionary sequences with low--mass secondaries 
($M_d $=1.25, 1.5 and 1.7$~M_\odot $) and black hole accretors 
($M_{bh} $=5 and 10$~M_\odot$) are calculated. The angular momentum 
losses due to magnetic braking and gravitational wave radiation 
are included. Using full nuclear networks (p--p and CNO cycles) we 
follow carefully the evolution of the surface composition of the 
secondary star. We find that the surface chemical composition of the 
secondary star may give additional information which
helps to understand the formation of  soft X--ray transients with 
black holes as accretors. We show that observations of isotope ratios 
$^{12}C/^{13}C$, $^{14}N/^{15}N$ and $^{16}O/^{17}O $ with comparison 
to computed sequences allow  estimates independent from spectroscopy of the 
mass of the secondary component.
We find that our evolutionary calculations  satisfactorily explain the 
observed
$\it q = M_{sg}/M_{bh}$ -- $P_{orb}$ distribution for Soft X--ray transients 
with orbital periods less than one day. Using our evolutionary 
calculations  we estimate secondary masses  and surface chemical
abundances (C,N,O) for different systems.  We distinguished three different 
phases in the SXT's evolution. The optical component shows (i) cosmic C, N, O 
abundances and $^{12}$C/$^{13}$C isotopic ratio; (ii) cosmic C, N, O
abundances but modified $^{12}$C/$^{13}$C ratio; and (iii) depletion of C and 
enhanced of N abundances and strongly modified isotopic ratios of C, N, O
elements.
\keywords{stars: evolution -- chemical evolution -- soft X--ray transients -- 
black holes}
}

\authorrunning{Ene Ergma and M.J. Sarna}
\titlerunning{What can we learn...}
\maketitle

\section{Introduction}

In recent years, about  a dozen of black hole candidates (BHC) with low--mass 
companion stars have been identified as the so--called soft X--ray transients 
(SXTs).
During the low X--ray luminosity stage (quiescent phase) optical/infrared 
observations of the optical companion allow the measurement of the mass 
function and sets a strong lower limit to the mass of the unseen companion

\begin{equation}
f(M_{bh})=\frac{M_{bh}^3 sin^3 \it i}{(M_{bh}+M_{sg})^2}=\frac{K_{sg}^3P_{orb}}
{2\pi G}
\end{equation}

where $M_{bh}$, $M_{sg}$  are the black hole and secondary masses, 
{\it i} is the inclination of the binary orbit, $P_{orb}$ is the orbital 
period and $K_{sg}$ is the velocity semi--amplitude of the optical component. 

If $ f(M_{bh}) > 3 
~M_\odot$, then $M_{bh}$ is larger than the upper limit to the gravitational 
mass of a  neutron star (Rhoades \& Ruffini 1974) and these compact objects 
are black holes. The total number of  close binaries with a black hole 
companion in the Galaxy is estimated to be between a few hundred and a 
few thousand. Only a small fraction of these systems shows X--ray activity 
(Tanaka \& Lewin 1995, Tanaka \& Shibazaki 1996).

The most popular scenario for the origin of low--mass X--ray binaries 
proposes as starting point a relatively wide binary system with extreme 
mass ratio (van den Heuvel 1983). After filling its Roche lobe, the massive 
primary engulfs its low--mass companion which will spiral--in inside its 
envelope (common envelope). In the common envelope scenario, if the 
envelope of the massive star is expelled before the 
low--mass secondary coalesces with the massive helium core of the primary 
then a close binary system forms.  Black hole formation occurs due to the 
collapse of massive helium core after the ejection of the red supergiant 
envelope.

The critical question is which stars end their evolution with black hole
formation.

For a long time it has been accepted that black holes form from very massive 
stars (more than 40--50$M_\odot$). Recently
 observational and theoretical evidence suggest that black holes form 
from stars with masses above 
25$~M_\odot$ 
(Portegies Zwart, Verbunt \& Ergma 1997, Ergma \& van den Heuvel 1998, 
Ergma \& Fedorova 1998, Fryer 1999).

The majority of observed black hole candidates (see Table 1) have orbital 
periods less than one day.
Therefore, the transient  nature of BHC systems and short orbital
periods put rather strong constraints on the properties of the progenitor 
systems and hence our understanding on how  these systems evolve. 

In this paper we would like to show that there is one additional, 
independent observational piece of evidence -- the abundance of CNO elements and their 
isotopic ratios --  
which will give us information about the progenitors of SXTs 
and their evolutionary stage.
 
\section{LMXB evolution -- general picture}

\begin{figure}
\centering
\includegraphics[width=8cm]{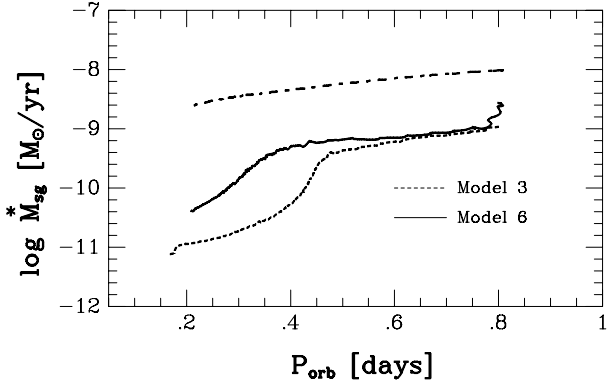}

\caption{The evolution of mass accretion rate as a function of the orbital 
period: dashed line -- model 3, solid line -- model 6. The critical mass 
accretion rate computed from Eq. (2) is also shown (dash--dotted line).}
\end{figure}
 
Following  the King, Kolb \& Burderi (1996) description we  discuss three cases 
of evolution depending on two timescales: 1) nuclear expansion of the 
secondary on the timescale $\tau_{ms}$ and 2) shrinkage of the orbit on the 
angular momentum loss timescale $\tau_{aml}$. In Case (1) $\tau_{ms}\ll
\tau_{aml}$, in Case (2) $\tau_{ms}\gg\tau_{aml}$, and in Case (3)  $\tau_{ms}\sim
\tau_{aml}$. For Case (1) the secondary evolves as a subgiant  (or giant) 
and transfers mass on the nuclear timescale.
The binary is evolving to longer periods and after mass transfer ceases 
a wide system with a black hole and a helium white dwarf is formed. A similar 
evolutionary scenario has been discussed for the formation of wide binary 
millisecond pulsars with helium white dwarfs (see recent results by 
Tauris \& Savonije 1999).  As shown by King et al. (1996) 
for Case (2)  the secondary is an unevolved main-sequence star and the mass transfer rate is always above the limit mass 
transfer rate (Equation (2)) for an irradiated disk, i.e. the mass 
transfer is stable.  The binary evolves to short periods (hours) and is observed as a  persistent X--ray source. In  Case (3), the secondary is evolved before mass transfer starts, but angular momentum losses shrink the binary orbit 
more rapidly than it expands and the binary evolves to very short periods 
(Ergma \& Fedorova 1998).

To understand why these sources are transients we use the dwarf nova instability criterion adapted to account for X-ray heating of the disk.
King, Kolb \& Szuszkiewicz (1997) have realized that heating 
by irradiation is 
much weaker if the accreting object is a black hole rather than a neutron 
star, since the black hole has no hard surface and cannot act as a point 
source for irradiation. For black hole binaries King et al. (1997) have 
obtained the following formula for the critical accretion rate

\begin{equation}
\dot{M}_{cr}^{irr}\approx 2.86\times 10^{-11}M_{bh}^{5/6}M_{sg}^{-1/6} 
P_{orb}^{4/3} ~~~~~[M_\odot~yr^{-1}]
\end{equation}

 For $\dot{M}$$<$
$\dot{M}_{cr}^{irr}$ mass transfer is unstable and the source will show 
transient outbursts.
  So we may exclude Case (2) in our consideration since for this case 
the mass transfer rate is stable and the binary is observed as a persistent 
but not a transient X--ray source.

Ergma \& Fedorova (1998) found the bifurcation period $P_{bif}$,  which 
separates  orbital evolution of Case (3) and  Case (2) from 
Case (1) to be about  
one day. So to have short orbital period systems it is 
necessary that the secondary  fills its Roche lobe  (RLOF) with 
 initial orbital period{\bf  $P_{i,orb}(RLOF)$} less
than $P_{bif}$. Also, the mass of secondary is not arbitrary but must be between 
1  and 1.8 $M_\odot$ to have  Case (2) or (3)  evolution. 
 Pylyser \& Savonije's (1988) calculations show an absence of  Case (2) 
or (3)
evolution for a binary consisting of $M_{bh}$= 4 $M_\odot$ with initial 
donor star mass $\geq$ 1.7 $M_\odot$.  

\begin{table*}
\caption[]{Observational data for SXT with known orbital periods ($<$ 1 d)}
\begin{center}
\begin{tabular}{llllll}
\hline
\multicolumn{6}{c}{}\\
Sources & $P_{orb}$ & \multicolumn{1}{c}{q} & \multicolumn{1}{c}{f($M_{bh}$)} 
& \multicolumn{1}{c}{i} & References\\
       & [d]     &    & \multicolumn{1}{c}{[$M_\odot$]} & &  \\
\hline
XTE J1118+480  & 0.171 &  &6.0$\pm$0.3& & McClintock et al. 2000, 2001\\
               &       & $\sim 0.05 $ &6.1$\pm$0.3& 81$\pm$2 & Wagner et al. 2001\\
GRO J0422+32   & 0.212 &$0.116^{+0.079}_{-0.071}$& 1.21$\pm$0.06 & 35--55 & 
Harlaftis et al. 1999\\
               &       &                         & 1.19$\pm$0.02 &         &
 Webb at al. 2000 \\
GRS 1009--45   & 0.285 & 0.137$\pm$0.015         & 3.17$\pm$0.12 & $<$80  &
Filippenko et al. 1999\\
A0620--00      & 0.323 & 0.067$\pm$0.010          & 3.18$\pm$0.16 & 31--54 &
Marsh et al. 1994\\
GS 2000+25     & 0.345 & 0.042$\pm$0.012          & 4.97$\pm$0.10 & 66    &
Casares et al. 1995, Harlaftis et al. 1996\\
GRS 1124--683  & 0.433 &$0.128^{+0.044}_{-0.039}$  & 3.10$\pm$0.4  & 54--65 &
Casares et al. 1997\\
H1705--250  & 0.521 &$<$0.053& 4.86$\pm$0.13   & 60--80 &
Harlaftis et al. 1997\\
\multicolumn{6}{c}{}\\
4U 1755-338      & 0.186 & \multicolumn{1}{c}{BHC} & & & Pan et al. 1995 \\     
XTE J1859+226  & 0.382 & \multicolumn{1}{c}{BHC} & & &Sanchez-Fernandez et al. 2000\\
GX339--4       & 0.617 & \multicolumn{1}{c}{BHC} & & & Tanaka \& Lewin 1995\\
\hline
\end{tabular}
\end{center}
\end{table*}

\section{The evolutionary code}

The Roche--filling--component (secondary star) models were computed
using a standard stellar evolution code based on the Henyey--type code
of Paczy\'nski (1970), which has been adapted to low--mass stars (Marks \& 
Sarna 1998). The
carbon--nitrogen--oxygen (CNO) tri--cycle affects the abundance ratios
we are interested in outside the hydrogen burning core. As the
secondary loses matter, due to mass transfer, layers
originally below the surface are exposed. As a consequence of mass
loss and nuclear evolution, a convective envelope develops and
penetrates to even deeper layers of the star, which are then mixed,
changing its surface chemical composition.

Our nuclear reaction network is based on that of Kudryashov \& Ergma (1980), 
who
included the reactions of the CNO tri--cycle in their calculations of
hydrogen and helium burning in the envelope of an accreting neutron
star . We have included the reactions
of the proton--proton (p--p) chain. Hence we are able to follow the
evolution of the elements  $^{1}$H, $^{3}$He, $^{4}$He, $^{7}$Be,
$^{12}$C, $^{13}$C, $^{13}$N, $^{14}$N, $^{15}$N, $^{14}$O, $^{15}$O,
$^{16}$O, $^{17}$O and $^{17}$F. We assume that the abundances of
$^{18}$O and $^{20}$Ne stay constant throughout the evolution. We use
the reaction rates of  Fowler, Caughlan \& Zimmerman (1967, 1975), Harris at 
al. (1983), Caughlan et al. (1985), Caughlan \& Fowler (1988), Bahcall \& 
Ulrich (1988), Bahcall \& Pinsonneault (1992), Bahcall, Pinsonneault \& 
Wasserburg (1995) and Pols et al. (1995).

\begin{table*}
\caption[]{The computed sequences}
\begin{center}
\begin{tabular}{lccccccc}
\hline
Model & $P_{i,orb}$ & $P_{orb}$ & $P_{orb}$ & $M_{i,bh}$ & $M_{i,sg}$ & 
$M_{sg}$ & $M_{sg}$\\
      & (RLOF)      & (C/N=1)   & (C/N=0.1) &            &            & 
(C/N=1)  & (C/N=0.1)\\
      & [d]      & [d]    & [d]    & [$M_\odot$]& [$M_\odot$]&
[$M_\odot$] & [$M_\odot$]  \\
\hline
1   & 0.6  &0.234 &0.191 &   10   & 1.25 & 0.472 & 0.346\\
2   & 0.7  &0.315 &0.277 &   10   & 1.25 & 0.488 & 0.365\\
3   & 0.8  &0.455 &0.431 &   10   & 1.25 & 0.514 & 0.353\\
4   & 0.7  &0.295 &0.249 &    5   & 1.25 & 0.487 & 0.372\\
5   & 0.7  &0.277 &0.309 &   10   & 1.50 & 0.624 & 0.505\\
6   & 0.8  &0.373 &0.336 &   10   & 1.50 & 0.636 & 0.505\\
7   & 0.9  &0.791 &0.599 &   10   & 1.50 & 0.671 & 0.599\\
8   & 0.8  &0.347 &0.312 &    5   & 1.50 & 0.632 & 0.507\\
9   & 0.8  &0.442 &0.394 &   10   & 1.70 & 0.749 & 0.610\\
10  & 0.8  &0.410 &0.355 &    5   & 1.70 & 0.746 & 0.611\\
\hline
\end{tabular}
\end{center}
\end{table*}

\begin{figure}
\centering
\includegraphics[width=8cm]{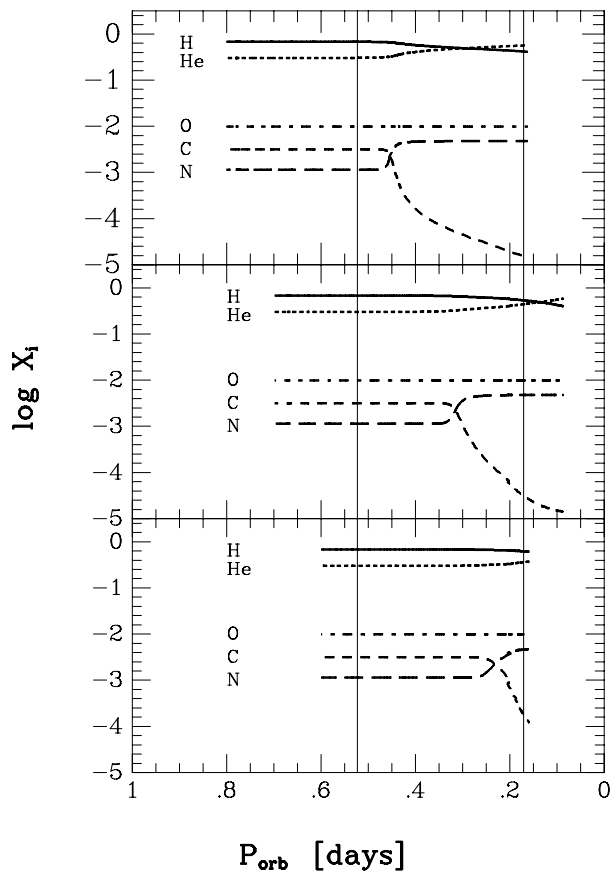}
\caption{The evolution of the red subgiant surface abundances of H, He, O, 
C, N  as a function of orbital
period: lower panel -- sequence (1), middle panel -- sequence (2), upper
panel -- sequence (3). The two thin solid vertical lines show the orbital period 
region for SXT's.}
\end{figure}

\begin{figure}
\centering
\includegraphics[width=8cm]{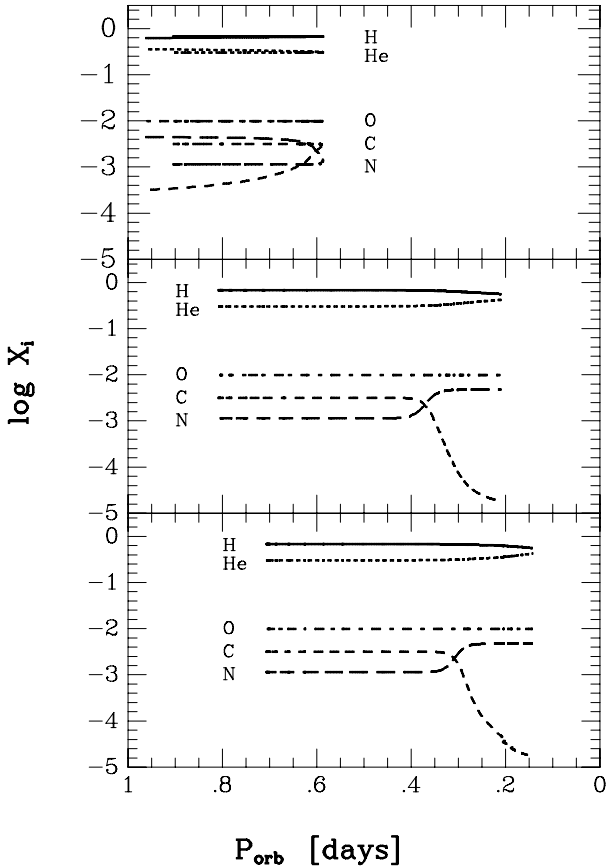}
\caption{The same as for Fig.2: lower panel-- sequences (5),
, middle panel -- sequences (6), upper panel--sequence (7)}
\end{figure}

For radiative transport, we use the opacity tables of
Iglesias \& Rogers (1996). Where the Iglesias \& Rogers (1996) tables are 
incomplete, we
have filled the gaps using the opacity tables of Huebner et al. (1977). For
temperatures less than 6000 K we use the opacities given by
Alexander \& Ferguson (1994) and Alexander (private communication). The
contribution from conduction, which is present in the Huebner et al. (1977) 
opacity tables, has been added to the other tables as well, since they
don't include it (Haensel, private communication).
The chemical  compositions X=0.7, Z=0.02.
In calculating evolutionary models of binary stars, we must take into
account mass transfer and the associated physical mechanisms which
lead to  angular momentum loss. 
We consider the subsequent mass
transfer from the secondary to the black hole when the
secondary star reaches contact with its Roche lobe. We use the
Eggleton (1983) formula to calculate the size of the secondary's
Roche lobe.
We take into account angular momentum losses due to gravitational wave 
radiation (Landau \& Lifshitz 1971) and magnetic stellar wind braking 
(Verbunt \& Zwaan 1981).

\section{Observational data}

In Table 1 we present observational data for all known or suspected SXTs 
with a black hole as accretor and orbital period less than one day.

Spectral observations and a mass function determination have been done 
for several systems. For a few systems  C, N, O spectral lines 
in different wavelength ranges were also detected.

\subsection{XTE J1118+480}

XTE J1118+480 is  BHC (Table 1) with an orbital period of 0.171 d (Patterson 
2000).  McClintock et al. (2000, 2001) found that f($M_{bh}$)=6.0$\pm$0.36 $M_
\odot$. Similar result has been presented by Wagner et al. (2001) (f($M_{bh}$)
=6.1$\pm$0.3 $M_\odot$). 
Haswell, Hynes \& King (2000) found that the Balmer jump appears in absorption. 
N V emission (124.0, 124.3 nm) is most prominent with equivalent 
width 0.6 nm. No C IV or O V emission is detected, suggesting that the 
accreting material has been CNO--processed.

\subsection{XTE J1859+226}

XTE J1859+226 is suspected BHC with suggested orbital period $P_{orb}$=0.382 d 
(Sanchez-Fernander et al. 2000). The ultraviolet spectrum shows broad and deep Ly--$\alpha$
absorption, strong C IV 155.5 nm emission 
and weaker emission lines of C III, N V ,O III, O IV, O V, Si IV and He II 
(Hynes et al. 1999).

\subsection{4U 1755--338}

The spectrum is nearly featureless. Very weak HeII was measured in one
spectrum in 1986 (Cowley, Hutchings \& Crampton 1988).

\subsection{GRS 1124--68}

Della Valle, Jarvis \& West (1991) 
 found that the most prominent emission lines are 
$H_{\alpha}$, $H_{\beta}$, N III+He II and N II (721.7 nm). The  
N III emission is normally attributed to the X-ray heating driving 
the Bowen fluorescence process, but in the present case it appears 
broadened by the O II and C III ions.

\section{Results of calculations}

 In Table 2 we present computed masses and orbital periods for two different
phases of the binary evolutionary sequences: for the beginning of carbon
depletion (C/N=1) and the high carbon depletion (C/N=0.1).

\begin{figure}
\centering
\includegraphics[width=8cm]{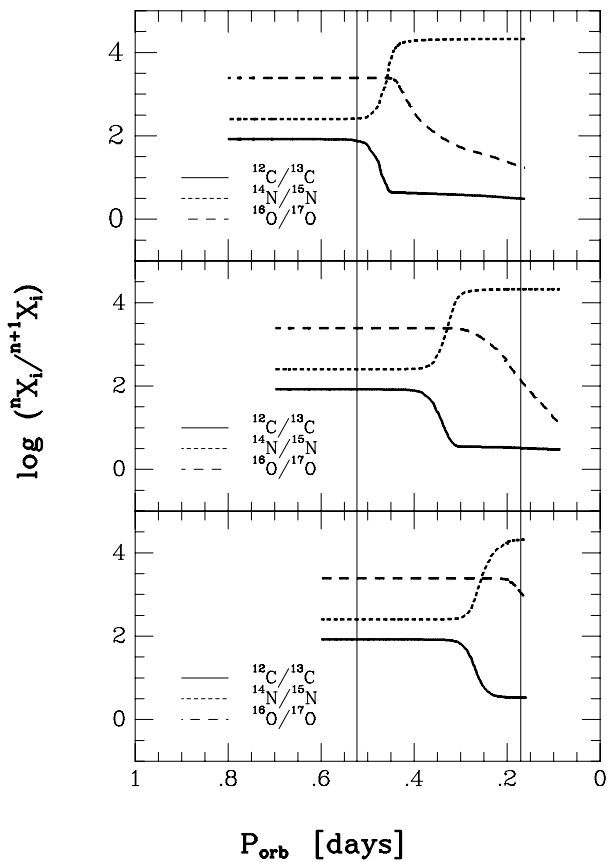}
\caption{ The evolution of various isotope ratios $^{12}$C/$^{13}$C, 
$^{14}$N/$^{15}$N and $^{16}$O/$^{17}$O versus orbital period for
1.25$~M_\odot $ secondary: lower panel -- sequence (1), middle panel -- 
sequence (2), upper panel -- sequence (3). The two thin solid vertical 
lines show the orbital period region for SXT's.}
\end{figure}

\begin{figure}
\centering
\includegraphics[width=8cm]{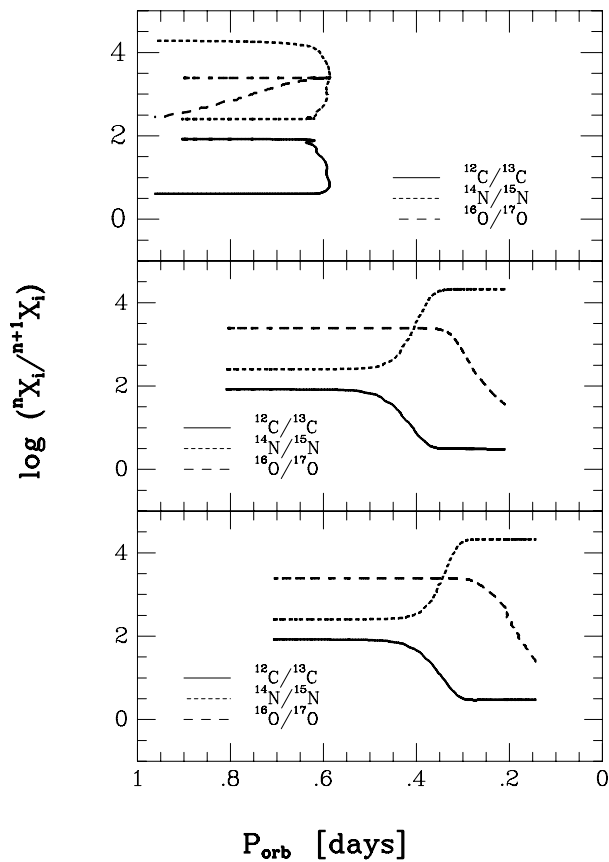}
\caption{The same as for Fig. 4 but for 1.5$~M_\odot$ secondary: lower 
panel -- sequence (5) ,middle panel -- sequence (6), upper panel -- 
sequence (7)}
\end{figure}

\begin{figure}
\centering
\includegraphics[width=8cm]{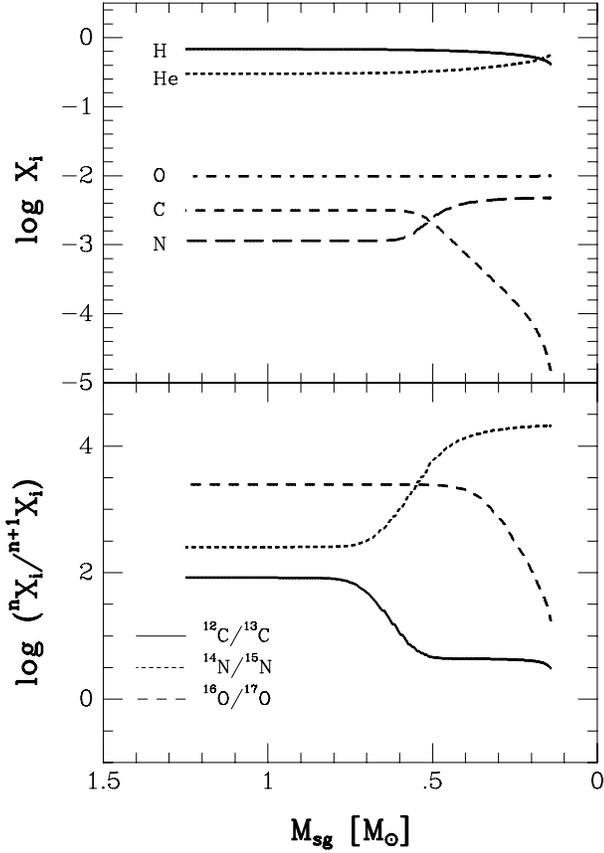}
\caption{The evolution of red subgiant surface abundances of H, He, O, C 
(upper panel) and isotopic ratios of  $^{12}$C/$^{13}$C, 
$^{14}$N/$^{15}$N and $^{16}$O/$^{17}$O (lower panel) as function of 
secondary mass (Model 3).}
\end{figure}

\begin{figure}
\centering
\includegraphics[width=8cm]{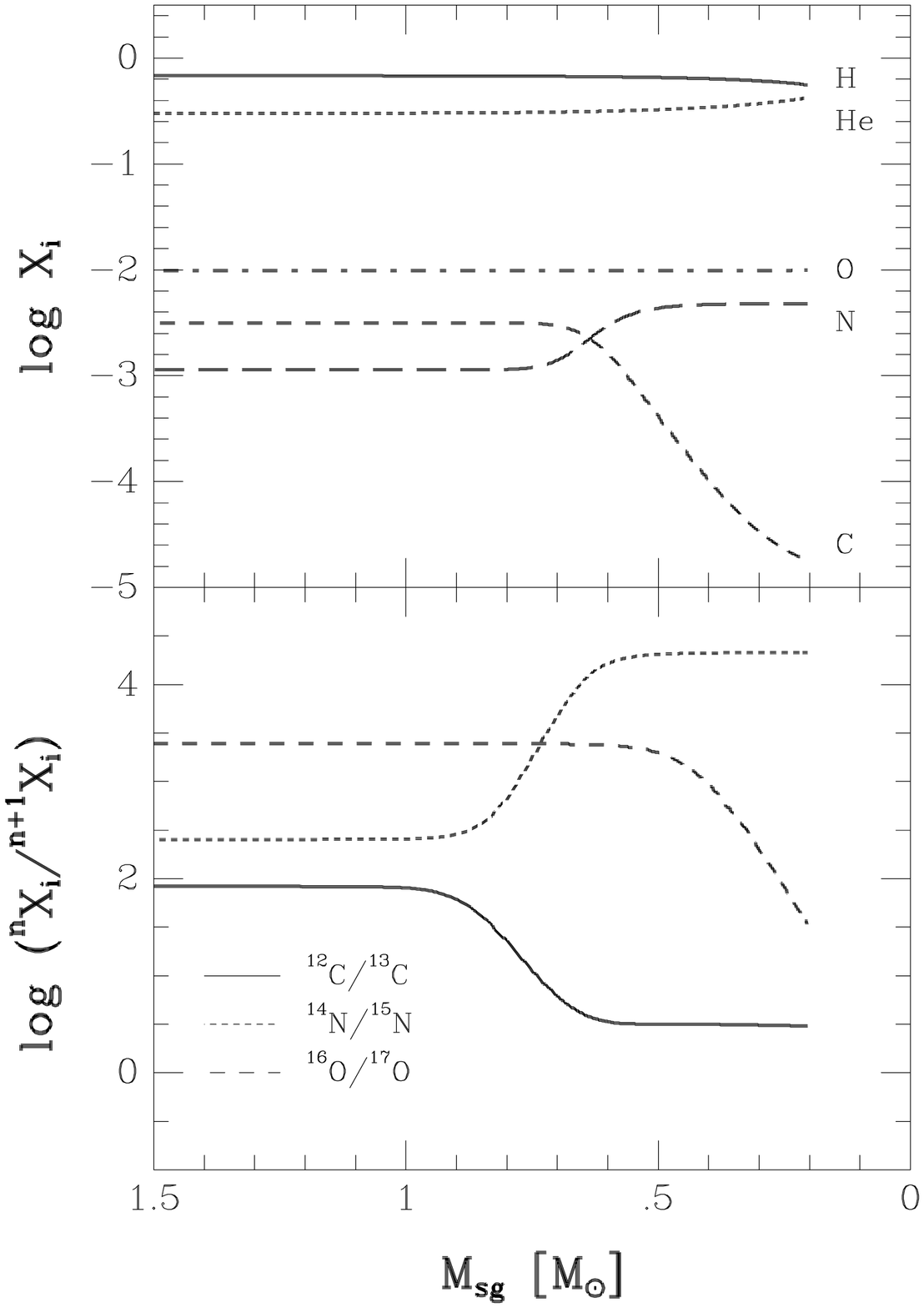}
\caption{The same as for Fig. 7 but for model 6.}
\end{figure}

 In our Table 2 only for Model 7  the secondary fills Roche lobe  
close to the bifurcation period and binary evolution ends up with almost the
same period as the initial orbital period. For other models,
since $P_{i,orb}(RLOF)$$<$$P_{bif}$ all binaries evolve towards short orbital 
periods. In Fig. 1 we present mass accretion rate versus orbital period for 
Models 3 and 6 . The critical mass accretion rate  (Equation (2)) is also 
shown (dash--dotted line). Our calculations show that during evolution the 
secular mass accretion rate is always less than the critical one.
Ergma \& Fedorova (1998) pointed out that the 
condition for  disk instability is more favorable in  systems with 
higher black hole masses. 

Our main interest was concerned with how the surface chemical composition 
evolves depending on the initial secondary mass and the initial orbital period 
of RLOF. In Fig. 2a, b, c we present the evolution of
the 1.25$~M_\odot $ red subgiant surface abundances  
of H, He, O, C, N as a function of orbital period for Models 1, 2 and 3. 
In Fig 3a, b, c the same is shown but for 1.5$~M_\odot $ 
secondaries (Models 5, 6, 7).

Figures 2 and 3 show that surface chemical composition may give us 
additional information about the progenitors of SXTs. If 
we do not observe carbon in the spectra of the  optical companion of SXTs 
with orbital period about 0.4 d,  then the secondary must fill its 
Roche lobe near $P_{i,orb}(RLOF)\geq$0.8 d (but less than one day). 
The secondary initial mass must be less than 1.7 $M_\odot$. 
We  predict that if we observe carbon in the spectra of the secondary 
star in SXT with $P_{orb} \simeq$0.3 d, then the
initial mass of the secondary must be less than 1.5 $M_\odot$ (Fig.3). 
For $M_{sg}=1.25~M_\odot$, initial Roche lobe filling must occur when 
$P_{i,orb}(RLOF)\leq$0.6 d. In Figs. 4a, b, c and 5a, b, c the
dependence of various isotope ratios $^{12}$C/$^{13}$C, $^{14}$N/$^{15}$N 
and $^{16}$O/$^{17}$O versus orbital period are shown.

\begin{figure}
\centering
\includegraphics[width=8cm]{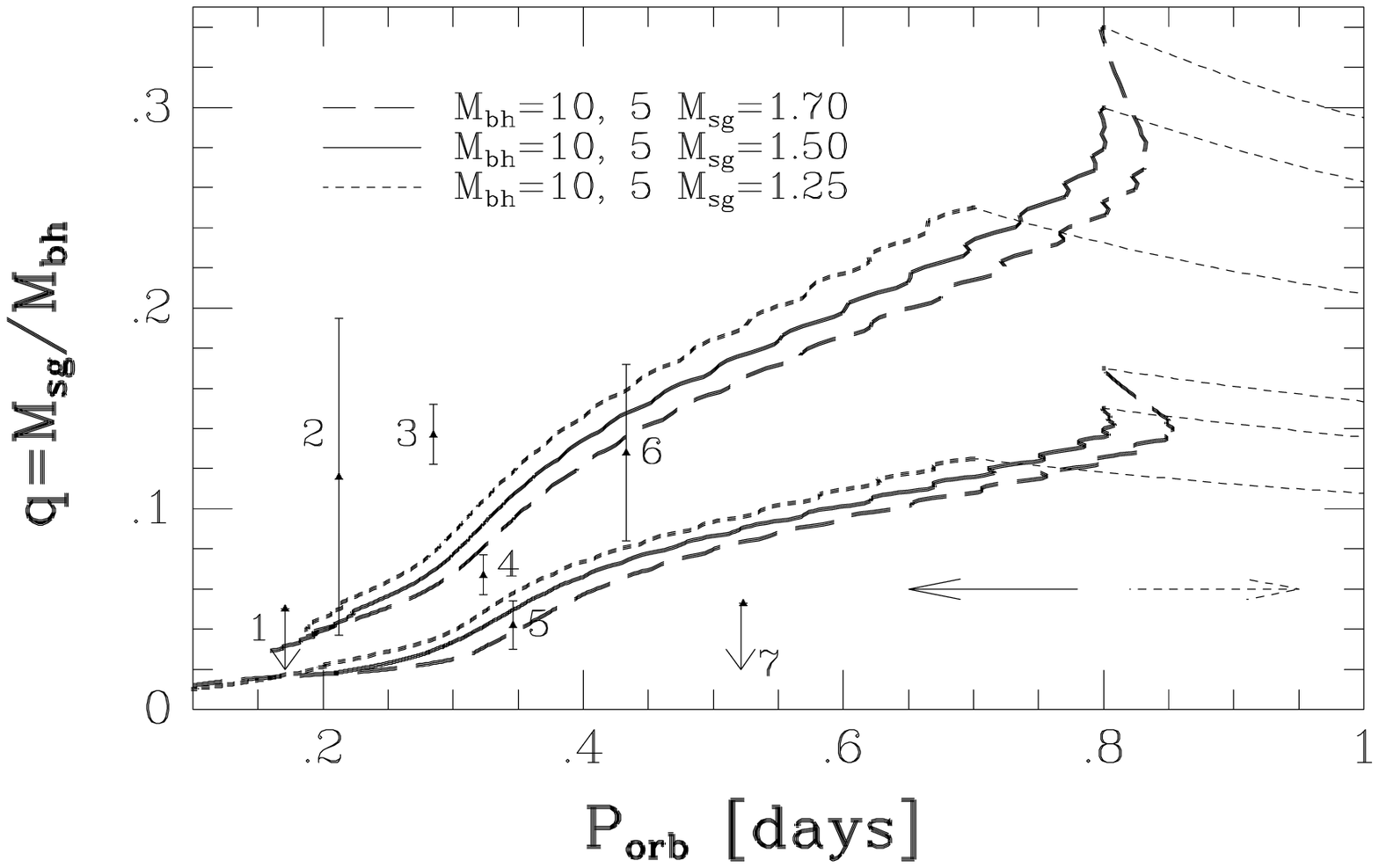}
\caption{The observed mass ratio {\it q} versus orbital period $P_{orb}$ for 
black hole X--ray transients: 1 -- XTE J1118+480, 2 -- GRO J0422+32, 3 -- GRS 1009--45, 
4 -- A0620--003, 5 -- GS 2000+25, 6 -- GRS 1124--683, 7 -- H1705--250. 
The calculated  theoretical relations ({\it q},$P_{orb}$)
for various $M_{bh}$ and $M_{sg}$ are also plotted.  Thin dashed lines show 
fully conservative evolution. The two arrows show the directions of
non--conservative (solid) and fully conservative (dashed) evolution. 
Upper curves are for $M_{bh}$= 5 $M_\odot$, lower for $M_{bh}$=10 $M_\odot$}
\end{figure}

\begin{figure}
\centering
\includegraphics[width=8cm]{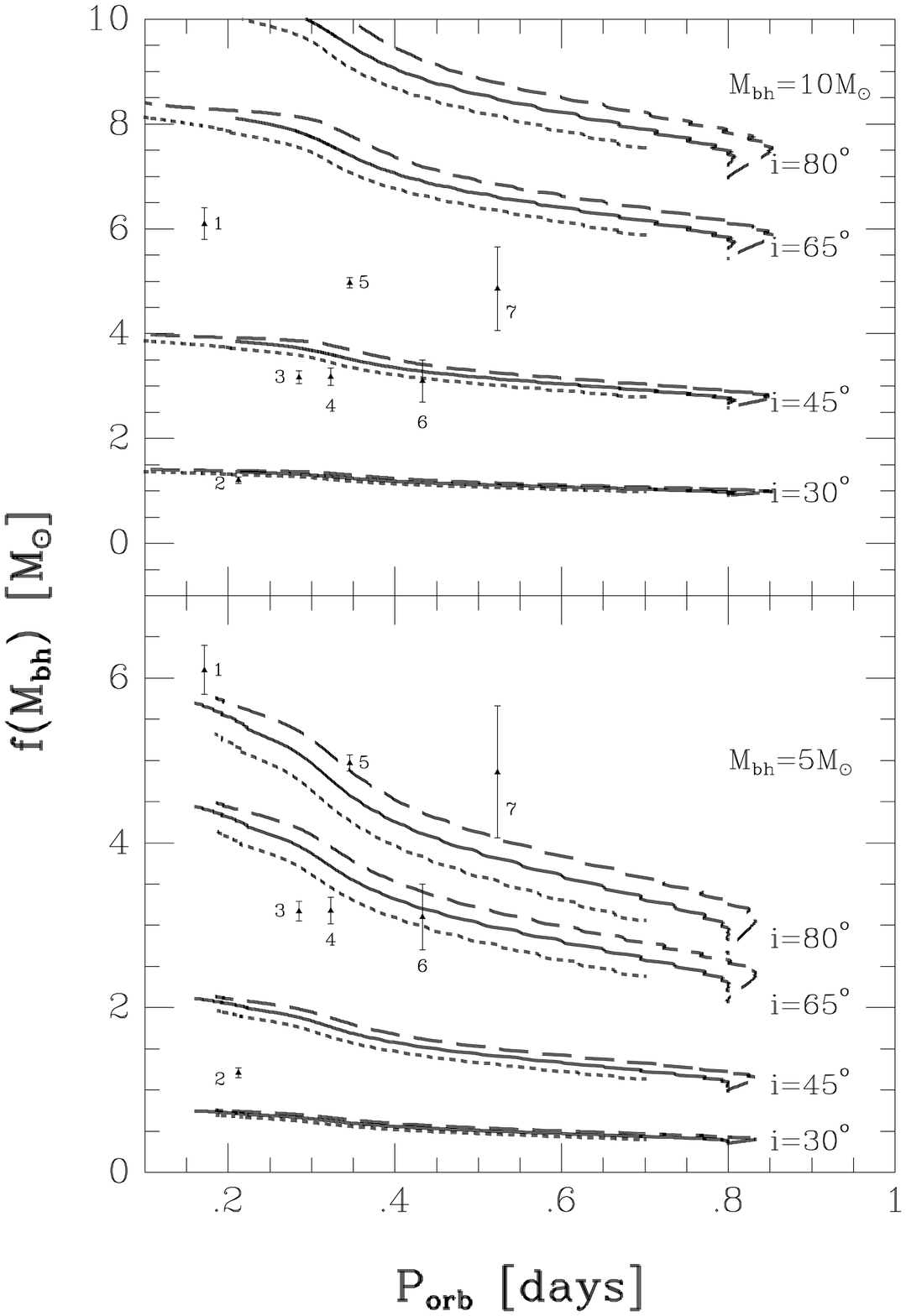}
\caption{The observed mass function f($M_{bh}$) versus orbital period 
$P_{orb}$ for 
black hole X--ray transients: 1 -- XTE J1118+480, 2 -- GRO J0422+32, 3 -- GRS 1009--45, 
4 -- A0620--003, 5 -- GS 2000+25, 6 -- GRS 1124--683, 7 -- H1705--250.
The calculated  theoretical relations ($f(M_{bh}$),$P_{orb}$)
are for $M_{bh}$=10$M_\odot$ and $M_{sg}$ =1.25 (short--dashed lines), 
1.5 (solid lines) and 1.7$M_\odot$ (long--dashed lines) (upper panel).  
{\it i} = 30, 45, 65, 80  are also marked. The same  for $M_{bh}$=5$M_\odot$ 
and $M_{sg}$ =1.25 (short--dashed lines), 1.5 (solid lines) and 
1.7$M_\odot$ (long--dashed lines) (lower panel).}
\end{figure}

In Fig. 6a, b we show the evolution of the red subgiant surface abundances 
of H, He, O, C, N (a) and 
$^{12}$C/$^{13}$C, $^{14}$N/$^{15}$N and $^{16}$O/$^{17}$O (b) as a 
function of secondary mass (Model 3). In Fig. 7a, b the same  
is shown for Model 6. From these figures we can see that for more 
massive secondaries isotope ratios start to change when the mass of the 
secondary has decreased to 0.9$M_\odot$ (0.7$M_\odot$ for a less massive 
secondary) and C/N ratios change when $M_{sg}\leq$0.6$M_\odot$ (for 1.5 
$M_\odot$) and $M_{sg}\leq$0.5$M_\odot$ (for 1.25 $M_\odot$) independent 
of the initial period of the RLOF. 

 From the point of view of the chemical composition evolution we will 
distinguish three different phases of the SXT's evolutionary stage:

\noindent
1) For $M_{sg} > ~0.7 ~M_\odot $ and $ P_{orb} > 
0.4 ~d$ the optical component shows solar/cosmic C, N, O abundances and 
isotopic ratio of $^{12}$C/$^{13}$C.

\noindent
2) For $0.4 ~M_\odot < M_{sg} < 0.7
~M_\odot $ and $ 0.3 ~d < P_{orb} < 0.4 ~d $ the optical component shows 
cosmic C, N, O abundances but modified $^{12}$C/$^{13}$C ratio.

\noindent
3) For $M_{sg} < ~0.4 ~M_\odot $ and $ P_{orb} < 0.3 ~d$ the optical 
component shows depletion of carbon and  enhanced of N abundances (small C/N
ratio) and strongly modified isotopic ratios of C, N, O elements.

We  conclude that chemical evolution 
can give us  extra information on the  mass of the secondary and this 
is independent of the black hole mass and the initial evolutionary stage of 
the secondary component (see Table 2).

\subsection{Comparison with observations}

In Fig. 8 we present calculated  {\it q} versus orbital period $P_{orb}$ 
 for $M_{sg}$=1.70, 1.50, 1.25 $M_\odot$ and $M_{bh}$= 5$M_\odot$ 
(upper curves) and $M_{bh}$=10 $M_\odot$ (lower curves).  
 In this figure we also plot observed {\it q} values from Table 1.  We see
that for five systems (XTE J1118+480, A0620--003, GS 2000+25, GRS 1124--683 
and H1705--250 (only upper limit)) our theoretical results agree 
satisfactorily with the observations. For GRO J0422+32 and 
GRS 1009--45 agreement is not good. The mass ratio for GRS 1009--45 is based 
only on an $H_{\alpha}$ emission radial velocity curve that is not exactly 
in antiphase with respect to the absorption line velocity curve. Therefore 
this mass ratio should be treated with   extreme caution (referee remark). 
The mass ratio for GRS 1009--45 is not reliable so that point in {\it q} --
$P_{orb}$ diagram can move down (good for theoretical models) or up 
(bad for our picture).
For XTE J1118+480 Wagner et al. (2001) found that modeling of the light curves
gives a small mass ratio ($\sim$ 0.05) which fits well with our model 
calculation results (see also Fig.8).

In Fig. 9 computed mass functions (for different values of orbital 
inclination)  are shown for two black hole masses (5 and 10$M_\odot$) and 
three secondary masses (1.25, 1.5 and 1.7$M_\odot $).

 From computed evolutionary sequences 
we  predict the chemical composition of the optical companion of the black 
hole. 
For all systems with known orbital period (see Table 1) we estimate (from our
grid of models) the following parameters: C/N, $^{12}C/^{13}C $,
$^{16}O/^{17}O $ and $M_{sg} $. The data are presented in Table 3. 
For all systems we  predict the 
range of secondary masses using only orbital period and our  grid of
models. 

 All our models (beside Model 7) predict that near orbital period
0.2--0.3 d 
carbon is depleted (GRO J0422+32 and XTE J1118+480). For XTE J1118+480 we 
estimate a secondary mass in the range 0.14--0.29$~M_\odot $ and a rather small
mass ratio (q$<$ 0.04).

If XTE J1859--058 has an orbital period  of 0.382 d and its UV 
spectrum shows 
carbon emission lines then  besides the Models 3,7,9 and 10 carbon is not 
depleted near this orbital period.

\begin{table*}
\caption[]{The predicted chemical composition and secondary masses of SXTs 
with  orbital periods $<$ 1 d}
\begin{center}
\begin{tabular}{llllll}
\hline
\multicolumn{6}{c}{}\\
Sources & \multicolumn{1}{c}{$P_{orb} $} & \multicolumn{1}{c}{C/N} & 
\multicolumn{1}{c}{$^{12}C/^{13}C$} & \multicolumn{1}{c}{$^{16}O/^{17}O$} & 
\multicolumn{1}{c}{$M_{sg}$} \\
       & \multicolumn{1}{c}{[d]}  &  &  &  & \multicolumn{1}{c}{[$M_\odot$]} \\
\hline
XTE J1118+480  & 0.171 & 0.005--0.04 & 3--3.4   & 20--1100 & 0.14--0.29\\
GRO J0422+32   & 0.212 & 0.004--0.02 & 3--3.4   & 30--500  & 0.15--0.36 \\
GRS 1009--45   & 0.285 & 0.01--1.14  & 3.1--3.8 & 50--2000 & 0.17--0.54 \\
A0620--00      & 0.323 & 0.005--1.6  & 3--5.2   & 70--2470 & 0.18--0.67 \\
GS 2000+25     & 0.345 & 0.02--2.78  & 10--83   & 100--2470& 0.20--0.83 \\
GRS 1124--683  & 0.433 & 0.12--2.78  & 4--83    & 1600--2470 & 0.36--1.05\\
H1705--250     & 0.521 & 2.78        & 74--83   & 2470     & 0.77--1.24 \\
\multicolumn{6}{c}{}\\
4U 1755--338   & 0.186 & 0.005--0.09 & 3--3.4   & 20--1800 & 0.15--0.33\\
XTE J1859+226  & 0.382 & 0.03--2.78  & 7--83   & 200--2470 & 0.25--0.90\\
GX339--4       & 0.617 & 2.78        & 83       & 2470     & 0.94--1.25\\
\hline
\end{tabular}
\end{center}
\end{table*}

\subsection{Observational tests}

Besides the UV and blue spectral regions it is interesting to try to observe
in red spectral regions (Charles 2001, private communication) and infrared 
regions of the spectrum.

\subsubsection{Red observations of the CN bands}

The red CN bands $A^2 \Pi - X^2 \Sigma^+$ (Bauschlicher, Langhoff \& Taylor
1988) from 4370--15050\AA ~~is useful for observations. We propose to
observe spectral region near 7920--7940\AA ~~ to identify two $^{13}$CN lines 
at 7921.13\AA ~~ and 7935.67\AA ~~ which are very useful for both 
$^{12}$C/$^{13}$C
isotopic ratio and C, N abundances determination. It revealed that in the
case of low $^{12}$C/$^{13}$C ratio ($<$ 10) these lines is clearly recognized
(Fujita 1985). However, in stars of high $^{12}$C/$^{13}$C ratio ($>$ 20) 
both isotopic lines is undetectable. The best candidates for such
observations are GRS 1009--45, A0620--00 and GRS 1124--683. The $E_{B-V} $ in
the direction of these systems is not very high and also optical companions 
are not too faint. For first two systems we predict no carbon detection 
while for last one carbon lines must be visible.

\subsubsection{Infrared observations of the CO bands}

Following Sarna et al. (1995), Marks, Sarna \& Prialnik (1997) and 
Marks \& Sarna (1998) the isotopic 
ratios $^{12}C/^{13}C$ and $^{16}O/^{17}O$ can be determined by infrared 
observations of the CO bands. Specifically, the bands of $^{12}CO$ and 
$^{13}CO$ around 1.59, 2.3 (2.29, 2.32, 2.35 and 2.38 $\mu$m) and 4.6 
$\mu$m (Bernat et al. 1979, Harris \& Lambert 1984a, b, Harris, Lambert \& 
Smith 1988) give a direct measurement of the isotopic ratio 
$^{12}C/^{13}C$. If we can estimate also the $^{16}O/^{17}O$
ratio then it is possible to determine the secondary mass using computed 
sequences. For such observations 10m class telescope is necessary.

\section{Conclusions}

Our theoretical models show that observations of chemical 
abundances may give additional information about the progenitors of SXTs and 
also about the mass of the secondary component. To produce the majority 
observed short orbital 
period SXTs with a black hole as accretor, the initial secondary mass must be 
between 1 and 1.7 $M_\odot$ and the initial orbital period (when the secondary is
filling its Roche lobe) between 0.5 and 1 d. It will be interesting to do 
population synthesis analyses to see how many systems it is possible to 
produce in the suggested orbital period -- secondary mass range.
Having data about observed C, N, O  and their isotopes abundances it is 
possible to estimate the mass of the secondary component.  
 Non--conservative evolution (in a sense of the orbital angular momentum 
loss from the  system) is able to explain satisfactorily the observed mass 
ratio and orbital period distributions. From our analysis one can conclude 
that the black hole masses are between 5 and 10 $M_\odot$  which agrees 
well with Bailyn et al. (1998)
results who compiled the observations of the mass functions and the best 
estimates of the mass ratios and inclinations and concluded that the 
black hole masses were clustered near 7 $M_\odot$. 

\section*{\sc Acknowledgements}
EE thanks Dr. D.Hannikainen for careful reading this paper and useful remarks.
EE acknowledges warm hospitality of the Astronomical Observatory Helsinki 
University where this paper was prepared. MJS thanks prof. P. Charles for
very useful discussions during his stay in University of Southampton. 
We thank anonymous referee for
his/hers very constructive referee opinion which really improve the text of 
this paper. This work was partially supported 
by a grant N 157992 from the Academy of Finland to Dr. Osmi Vilhu 
and ESF grant N 4338.
At Warsaw, this work has been supported through grants 2--P03D--014--13 and 
2--P03D--005--16 by the Polish National Committee for Scientific Research and
by the NATO Collaborative Linkage Grant PST.CLG.977383.

\end{document}